\font\tenbg=cmmib10 at 10pt
\def \rvecmu{{\hbox{\tenbg\char'026}}}

\def \Omegabold {{\hbox {\tenbg\char'012}}}

\documentstyle[emulapj,epsfig]{article}
\begin{document}

\bigskip

\font\two=cmbx10 scaled \magstep1

\title{\bf Magnetospheric Gap and Accumulation of Giant Planets
 Close  to the Star}

\author{M.M.~Romanova}
\affil{Department of Astronomy, Cornell University, Ithaca, NY
14853-6801; ~ romanova@astro.cornell.edu}

\author{R.V.E.~Lovelace}
\affil{Departments of Astronomy and Applied Physics, Cornell
University, Ithaca, NY 14853-6801; ~RVL1@cornell.edu }


\begin{abstract}

The bunching of giant planets at a  distance of several stellar
radii may be explained by the disruption of the inner part of the
disk by the magnetosphere of the star during the T Tauri stage of
evolution. The rotating magnetic
field of the star gives rise to a low
density magnetospheric gap where
stellar migration is strongly suppressed.
   We performed full 3D magnetohydrodynamic simulations of the
disk-magnetosphere interaction and examined conditions for which
the magnetospheric gap is ``empty", by changing the misalignment
angle between magnetic and rotational axes of the star, $\Theta$,
and by lowering the adiabatic index $\gamma$, which mocks up the
effect of heat conductivity and cooling.
    Our simulations  show that for a wide range
of plausible conditions the gap is essentially empty.
  However, in  the case
of  large misalignment angles $\Theta$,
part of the funnel stream is located
in the equatorial plane and the
gap is not empty.
   Furthermore, if the adiabatic index is small
($\gamma \sim 1.1$) and the rotational and magnetic axes are almost
aligned, then matter penetrates through the magnetosphere
due to 3D instabilities
forming high-density equatorial funnels.
  For these two limits there is appreciable matter
density  in the equatorial plane of the disk so that a planet may
migrate into the star.

\end{abstract}

\section{Introduction}

   More than  $170$  giant planets have been
discovered around solar-type stars (see catalog of planets in
http://vo.obspm.fr/exoplanetes/encyclo/catalog-main.php).
  About $24\%$ of them are located very close to the star,
at $r \lesssim  0.1 {\rm AU}$ (see also reviews by Marcy et al.
2003; Papaloizou \& Terquem 2006). There is a prominent peak in the
radial distribution of the planets at $r\sim 0.04 - 0.05~ {\rm AU}
\approx (7-10) R_\odot$ which corresponds to periods of 3 days (see
Figure 1).

  The radial distribution and other properties
of the giant planets have stimulated work  on  models of planet
formation and migration.
    According to the presently favored  interpretation,
planets form far away from the star either through  core
accretion (Mizuno 1980; Pollack  et al. 1996), or through
instabilities in the disk (Boss 2001).
   Subsequently they migrate inward
due to their gravitational interaction
with the disk (Lin \& Paploizou 1986;
Lin, Bodenheimer \& Richardson
1996; Ward 1997; Nelson \& Papaloizou 2003).

  Some planets may migrate close to the star
where the disk properties are strongly
influenced by the star and/or
the star's magnetic field.
   In particular, the inner regions of the disk
may be truncated as a result of
heating by the star (e.g., Kuchner \&
Lecar 2002) and
photoevaporation of the region of the disk
$\lesssim 1$ AU
(Matsuyama, Johnstone \& Murray 2003).
  Or, the inner part of the disk may be disrupted by  the strong
magnetic field of the protostar (Lin et al.
1996).

      There are strong
observational arguments that young solar-type stars (T Tauri type
stars) have  strong magnetic fields.
  They are thousands of times
brighter than the Sun in the X-ray
(see review by Feigelson \&
Montmerle 1999) which is a sign
of their high magnetic activity.
    In a number of cases, direct
measurements of the Zeeman broadening imply a
$B$ field  (averaged over the surface of the star)
of the order of several kilo-Gauss, which is much stronger
than that the average field on the
surface of the  Sun (e.g., Basri et
al. 1992; Johns-Krull et al. 1999).
  The strongest magnetic fields are
probably associated with the
multipolar component (e.g. Safier 1998;
Johns-Krull et al. 1999; Smirnov et al. 2003).
   However, a significant
dipole component is also expected.
  It gives many observational
signs of magnetospheric accretion in the T Tauri stage (see review
by Bouvier et al. 2006) and also is observed in some circular
polarization measurements  (e.g., Valenti \& Johns-Krull 2004;
Symington et al. 2005).

A sufficiently strong dipole magnetic
field will truncate the disk at
a distance of several stellar radii forming a low-density
magnetospheric gap as shown in Figure 2.
   The migration rate
of planets in this gap will be greatly reduced.

 This work  analyzes the conditions where a low-density gap exists
between the inner edge of the disk and the
star's surface.
   Our analysis is based on 3D MHD simulations.
In particular, we investigate properties of the gaps
for different misalignment angles
$\Theta$ between the rotation axis of the star $\Omegabold_*$ and its
magnetic moment $\rvecmu$.
  We investigate conditions where part of the funnel
stream is in the equatorial plane close to the star.
   This  may
give an appreciable matter density in the magnetospheric
gap.
    Also we consider the possibility of
direct equatorial accretion from
the disk to the star due to 3D
instabilities.

\begin{figure*}[t]
\epsscale{1.1} \plotone{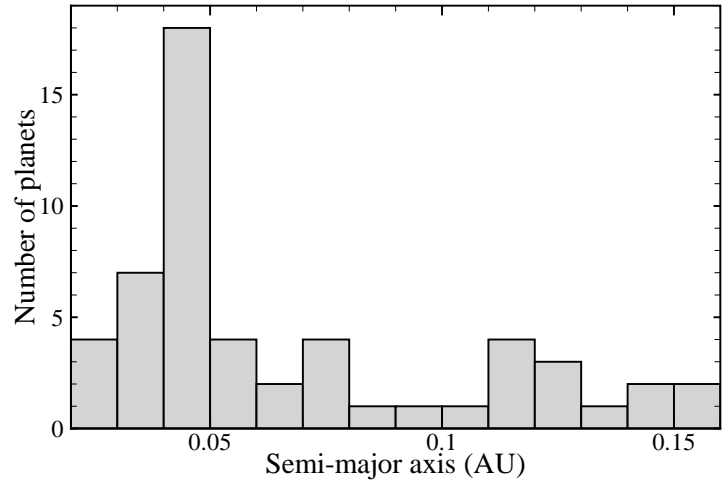}
 \caption{Distribution of extrasolar planets in the  vicinity of
the star.} \label{Figure 1}
\end{figure*}


\begin{figure*}[t]
\epsscale{1.3} \plotone{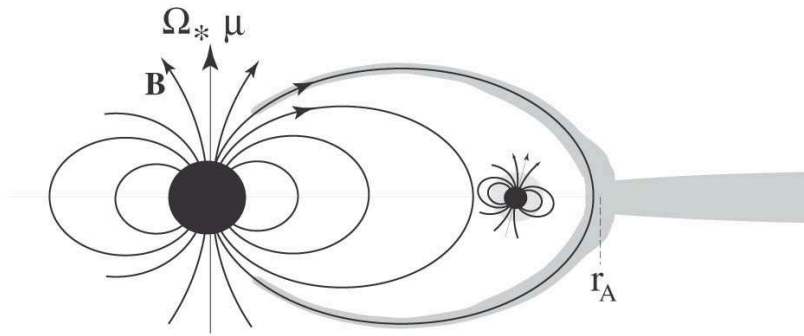}
 \caption{Sketch of an accretion  disk which is
disrupted by the star's dipole magnetic field.
  The rate of migration of a planet is greatly
slowed  once it enters the gap.}
\label{Figure 2}
\end{figure*}

\begin{figure*}[t]
\epsscale{0.9}
\plotone{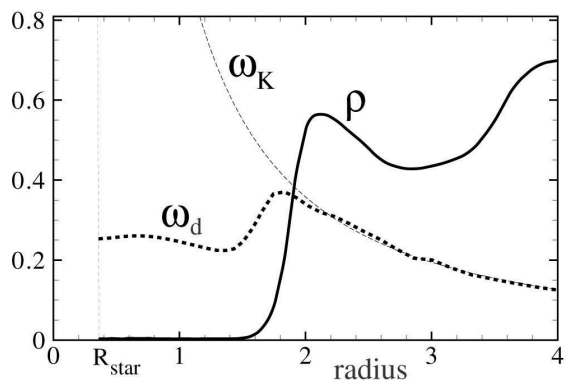}
\caption{Radial distribution of the
density (solid line) and the angular velocity of the disk
$\omega_d=v_\phi/r$ (dashed line) in the vicinity of the star for a
misalignment angle of $\Theta=30^\circ$. Thin dashed line shows
Keplerian angular velocity $\omega_K$.}
\label{Figure 3}
\end{figure*}

\begin{figure*}[t]
\epsscale{1.6} \plotone{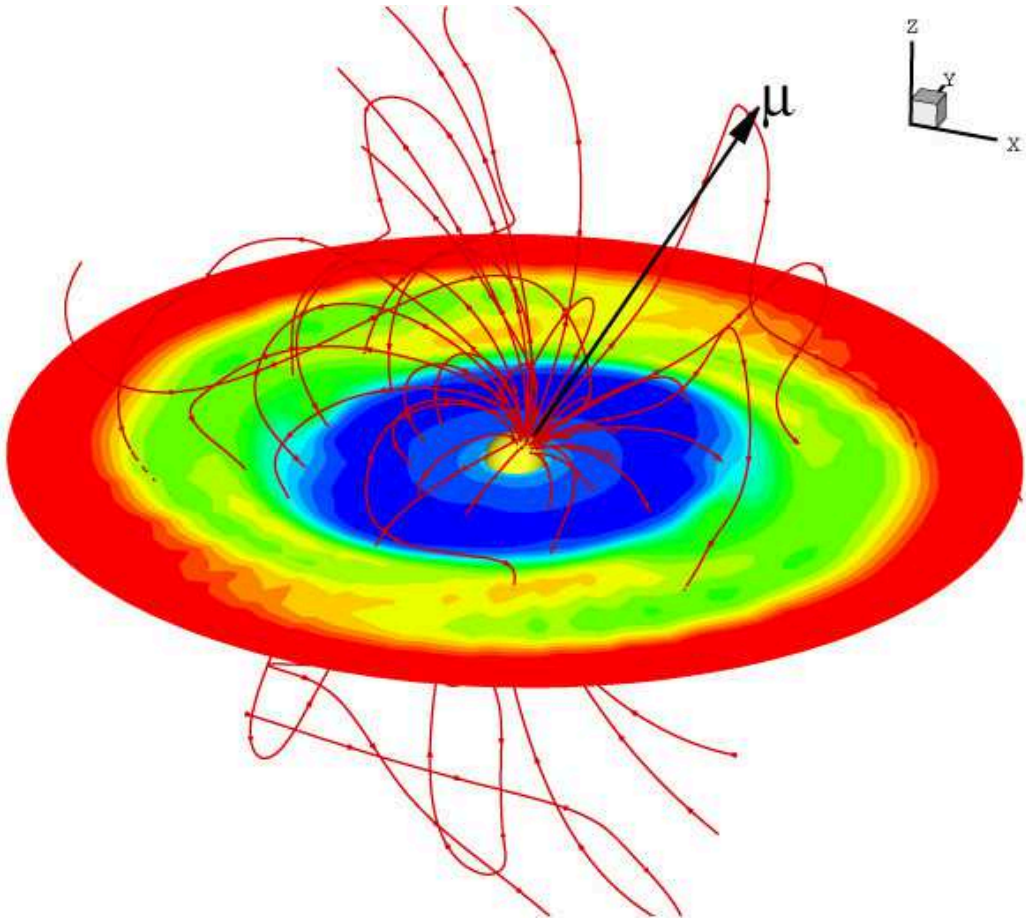} \caption{Result of 3D simulations of
disk accretion to a rotating star with a dipole moment $\rvecmu$
misaligned with the star's rotation axis $\Omegabold_*$ by
$\Theta=30^\circ$.
   The color background shows the density
distribution in the equatorial region.
 Density varies from $\rho\approx 0.003$
(blue color) to $\rho\approx 1$ (red
color).
  The red lines are magnetic field lines.
The black arrow shows the direction of the magnetic moment
$\rvecmu$.}
\label{Figure 4}
\end{figure*}

\begin{figure*}[t]
\epsscale{1.8} \plotone{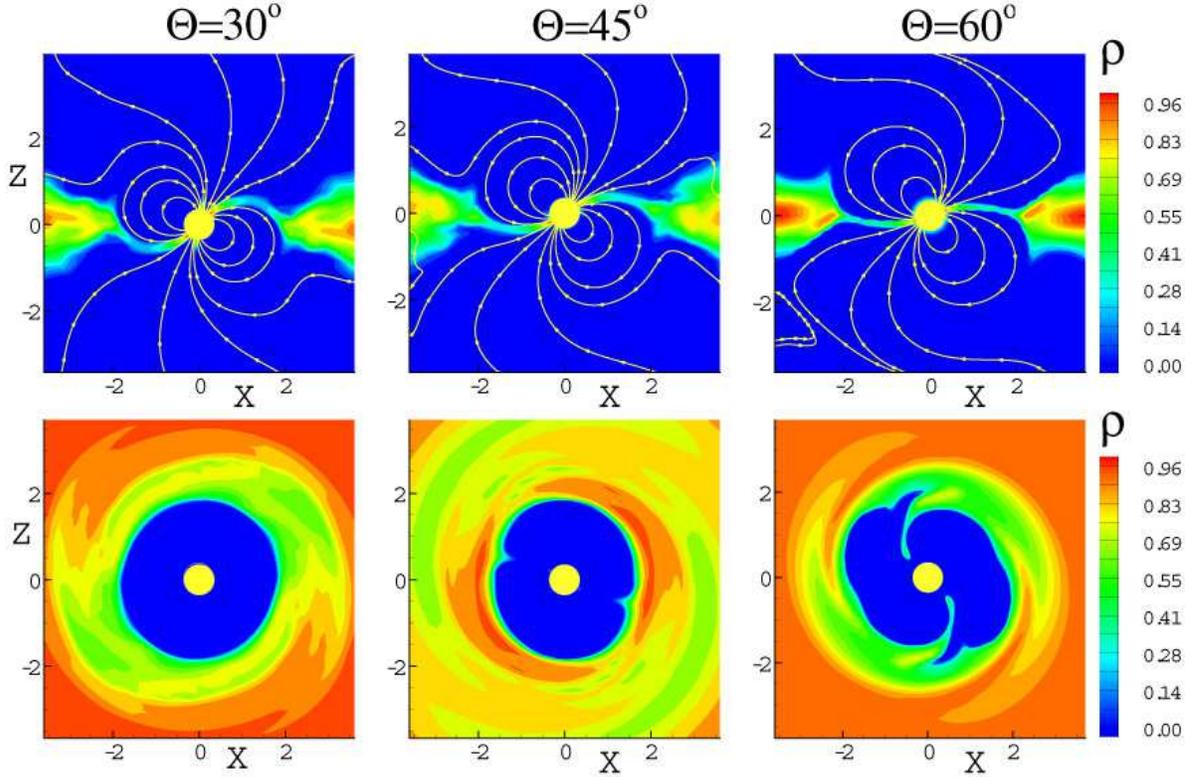} \caption{The top panels show the
density distributions and sample magnetic field lines in the
$\Omegabold_*-\rvecmu$ plane for different misalignment angles
$\Theta = 30^\circ, 45^\circ, 60^\circ$.
   The bottom panels show the equatorial distribution of
density.}
\label{Figure 5}
\end{figure*}

\begin{figure*}[t]
\epsscale{1.7} \plotone{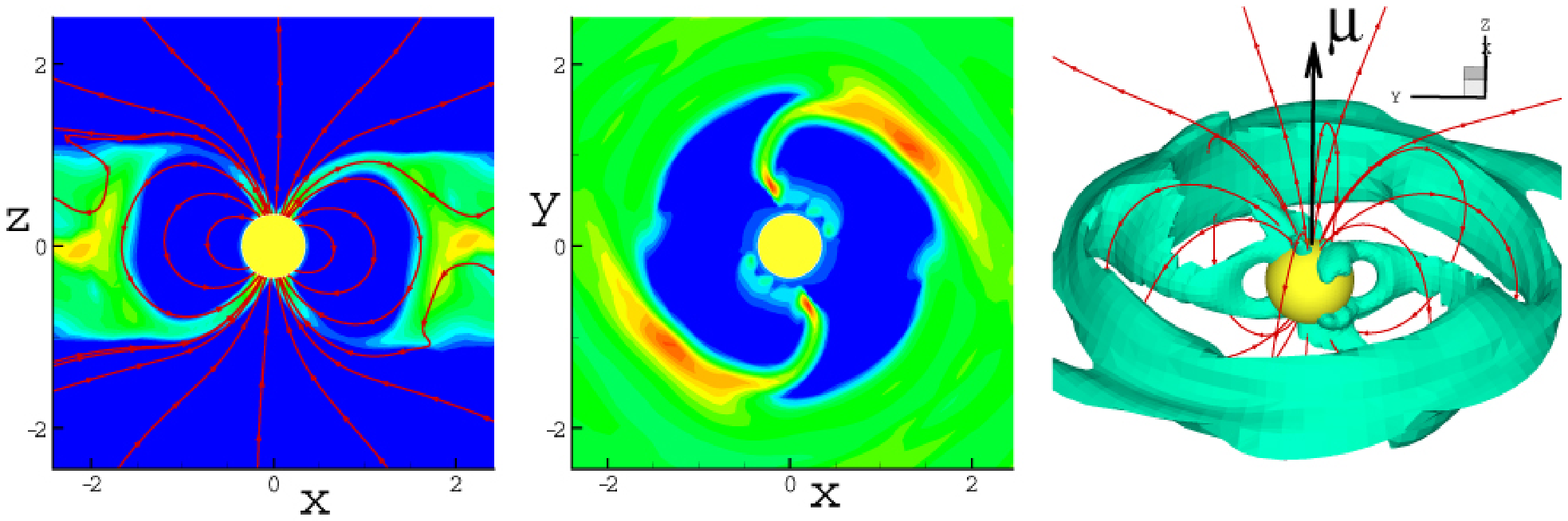} \caption{Result of simulations for
$\gamma=1.1$ for $\Theta = 5^\circ$ after $P=12$ rotations.
  The left panel
shows the density distribution
and sample field lines in the
$\Omegabold_* - \rvecmu$  plane.
  The middle panel shows the density distribution in the
equatorial plane, and the right
panel shows a 3D view  of the density levels and
sample field lines. }
\label{Figure 6}
\end{figure*}

\section{Where Does the Planet Migration Stop?}

The accretion disk is disrupted at the
Alfv\'en radius
where the dynamic pressure  of the
disk matter is comparable to the
magnetic pressure in the star's dipole field,
$r_A=[\mu^2/(\dot{M}\sqrt{GM})]^{2/7}$ or
\begin{equation}
r_A \approx 7.2\times 10^{11}~ {{B_3^{4/7} R_{2.5}^{12/7}}
\over{M_{0.8}^{1/7} \dot M_{-7}^{2/7}}}~{\rm cm}\approx 0.05~{\rm
AU}~,
\end{equation}
where  $M_{0.8}\equiv M_*/{\rm 0.8 M_\odot}$, $R_{2.5}\equiv
R_*/{2.5~R_\odot}$, and $B_3\equiv B_*/{10^3~{\rm G}}$ are the
normalized mass, radius, and magnetic field of the protostar, and
$\dot M_{-7}\equiv \dot M/{10^{-7}~{\rm M_\odot/{\rm yr}}} $ is the
accretion rate in the disk
(Ghosh \& Lamb 1979; Camenzind 1990;
K\"onigl 1991).
   For the typical parameters of T Tauri stars used in
this formula, this radius  coincides
approximately with the peak of
the distribution shown at Figure 1.
   Thus this peak may result from
the greatly reduced rate of migration inside
the magnetospheric gap.

Numerical simulations show that the disk is disrupted at the
distance $r\approx r_A$, where
  the plasma is
lifted out of the disk plane by the vertical
 pressure force and
it then flows along the star's dipole field lines in a funnel flow
(Romanova et al. 2002; 2003; 2004).
   As a consequence, the density of matter
in the equatorial plane is greatly reduced for $r < r_A$.
  Figure 3 shows the
equatorial density distribution obtained from our 3D
simulations.
  The density is large in the disk, and it often increases
as $r_A$ is approached.
However, for $r<r_A$ the density drops by
a factor of $\sim 100-300$
in the magnetically-dominated magnetosphere.
  A protoplanet which migrates inward to radii  $<r_A$
enters a region of greatly reduced density.

  For typical conditions  planets migrate inward as
a result of interaction of the planet with the disk matter. The
planet loses part of its orbital angular momentum by overtaking
collisions with the disk outside its orbit and it gains a smaller
part by overtaking collisions of the disk matter inside its orbit.
The rate of migration, or radial speed, $V_{pr}$, depends on a number
of parameters such as  mass of the planet, $M_p$, surface density of
the disk, $\Sigma$, viscosity in the disk, $\nu$, and others, and
also it is different in cases when a planet opens a gap in the disk
or not.

          If the planet's mass is
relatively small ($M_p\lesssim 10
M_\oplus$) it
does not open a gap in the accretion disk.
  The migration in this
case is referred to as ``type I,''    and the planet's inward drift
speed is $V_{pr} \propto -M_p (\Sigma r^2)$ (Ward 1997; Papaloizou
\& Terquem 2006).

           Planets of sufficiently large mass
open a gap in the disk of
width of the order of the disk thickness.
  The migration in
this case is referred to as ``type II,''  and it tends to ``lock''
the planet's migration to that of the disk matter if
the local disk mass, $M_d =4\pi r^2 \Sigma$ ,
is larger than the planets mass $M_p$.
  The disk matter moves inward with a
radial speed $v_r = - 3 \nu/(2r)$, where $\nu =
 \alpha c_s^2/ \omega_K$ is the usual Shakura-Sunyaev
turbulent viscosity with $\alpha = 10^{-3}-10^{-2}$.
    However, if the local disk mass is small compared
with the planet's mass, then the planets migration is
slower than that of the disk matter.
   The angular momentum lost by the planet
$d{J}_p/dt =M_p v_K V_{pr}/2$ in a second is equal to
the angular momentum transported outward
by the viscous stress in the disk, $d{J}_d/dt = \dot{M} r v_K$,
where $v_K=(GM/r)^{1/2}$ is the Keplerian velocity
and $\dot{M}$ mass accretion rate of the disk.
  This gives a migration speed of the planet
$V_{pr} = -(M_d/M_p)|v_r|$.
    We can write $\dot{M}=2\pi r \Sigma |v_r|$  so
that $M_d =4\pi r^2 \Sigma \approx 1.3 \times 10^{29}{\rm g}
(r/0.1{\rm AU})^{3/2}(\dot{M}/10^{-7}M_\odot/{\rm yr})$ for $h/r =0.1$ and
$\alpha = 10^{-3}$.

      For $M_d < M_p$,the time scale
for the planet's migration at
$r$ is simply
\begin{equation}
\tau_p = {r \over V_{pr} }= {M_p\over M_d}{2r^2 \over 3\nu}~,
\end{equation}
which is independent of $r$ and inversely
proportional to both $\dot{M}$
and $\alpha$.
  For a Jupiter mass planet, $\tau_p \approx 4700{\rm yr}
(10^{-7}M_\odot/{\rm yr}/\dot{M})$ for $\alpha =10^{-3}$ and $h/r=0.1$.
      However, equation(2) does not include the influence of
the star's magnetic field.
    Inside the magnetospheric gap the matter density is reduced
by a factor $\gtrsim 10^2$.
   The migration time will be increased by a corresponding
factor.

  Figure 3 also shows that the angular velocity of
the disk plasma in the equatorial plane $\omega_d$ within the
magnetospheric gap is much lower
than the Keplerian angular velocity of the
planet $\omega_K$ for stars with periods $P\gtrsim 2$ days.
  Thus, the disk
matter passing close by the planet will have a large relative
velocity $r(\omega_K-\omega_d)$.
  The relative velocity  is larger than in the non-magnetic
case by a factor $F\sim (r/h)(\omega_K-\omega_d)/\omega_K\gg1$ for
the type II migration.
    One can readily show that the rate of exchange of
angular momentum between the planet and the
disk is reduced by a factor $1/F^2 \ll 1 $.
    Thus the migration time inside the magnetospheric
gap will be increased further by a factor $F^2$.
We estimate $F^2 \gtrsim 10$.
   Once the planet is inside the magnetospheric gap
it may continue to migrate slowly inward owing to resonances
it has with the disk matter at larger radii but a
treatment of this is beyond the scope of the present
work.

Thus, in both cases the migration
 will be greatly reduced  if the
density in magnetospheric gap is
much lower than in the disk.
Conditions may be changed however
if the magnetic axis $\rvecmu$ is
misaligned relative to the rotational axis of the star
$\Omegabold_*$ by an angle $\Theta$.
   For high misalignment
angles $\Theta$, the funnel streams
may be partially located  in the
equatorial region (Romanova et al. 2003).
  From the other side, even in
the aligned case, some matter may accrete to the star in the
equatorial plane due to instabilities (e.g., Arons \& Lea 1976).
  We performed 3D simulations to investigate both factors.

\section{3D Simulations of the Disk-Magnetosphere Interaction
and Magnetospheric Gaps}

We performed a set of 3D simulations
using our code based on the
``cubed sphere" grid  (Koldoba et al. 2002) with the main goal
of analyzing
the density distribution in the magnetospheric gap.
  Simulations were set up in
a way similar to those of Romanova et
al. (2003, 2004).
  Namely,  quasi-stationary initial conditions
were used which permitted slow viscous
accretion from the disk to a star.
An $\alpha-$viscosity  was incorporated to the code with
typical values of $\alpha-$parameter: $\alpha=0.02$ and $0.04$.
The magnetic axis $\rvecmu$ is
misaligned relative to rotational axis
of the star $\Omegabold_*$ by an angle $\Theta$.
   The rotational axis of the
star coincides with that of the disk.

Simulations were done for parameters
typical for T Tauri type stars:
 $M_*=0.8 M_\odot$, $R_*= 2.5 R_\odot$,
$B_*=10^3 G$, $\dot M\approx 3\times 10^{-8}
{M_\odot/{\rm yr}}$.
   Compared to our previous 3D runs,
we changed parameters so as to
increase the size of magnetospheric
gap to $r_A=(4-5) R_*$ versus
$r_A=(2-3) R_*$ in our previous papers.
  Figure 4 shows the magnetospheric gap in a
test case with an  even larger
magnetosphere $r_A=(6-7) R_*$.
   One can see that the low-density
magnetospheric gap
can be quite large.
   One can also see that inside the
magnetospheric radius (which
corresponds approximately to the edge of the disk),
the magnetic field
lines are closed, while outside
of this radius they are carried by
the matter of the disk or corona.
 Our simulations show that the density
inside the magnetospheric gap is about
$100-300$ times smaller than
the density in the nearby disk.

\noindent{\bf Magnetospheric Gap at Different $\Theta$.} We
performed 3D simulations for different misalignment angles $\Theta$
from $\Theta=0^\circ$ to $90^\circ$.
  We investigated the
magnetospheric gaps in the equatorial plane. Simulations have shown
that matter flow is different at small and large misalignment
angles.
  For  angles $\Theta\lesssim 45^\circ$, matter
flows to the star along  funnel streams which are
above and below the equatorial plane.
   Thus within the magnetospheric gap $r<r_A$ the
matter density in the equatorial plane is greatly
reduced (see Figure 5, left two panels).
    For larger angles, matter also
accretes to a star through the funnel streams.
    However,  part of the funnel
streams is  located in the equatorial plane
and the magnetospheric gap
is not empty (Figure 5, right panel).
  Thus at large $\Theta$, the
planets orbiting in the equatorial
plane will interact with the
dense gas of the streams and may
continue to migrate inward to the
star.

\noindent{\bf Accretion through Equatorial Funnels at Low $\gamma$}.
  There is another possible reason why
the magnetospheric gap may  have some matter density.
   There are possible instabilities which may
lead to the direct accretion of
matter through the magnetosphere
in the equatorial plane.
   To investigate such instabilities,
we took the almost aligned case,
$\Theta=5^\circ$, and decreased
the adiabatic index from $\gamma=5/3$ to $\gamma=1.1$.
  The adiabatic index may be significantly lower
than its  ideal value in the case of  high electron
heat conductivity which may occur in a highly
ionized plasma.
  In  our simulations, the low value of $\gamma$
acts to give a low temperature
in the disk and the funnel flow.
  We observed that  matter partially accretes in
the equatorial plane.
   Figure 6 shows that matter accreted
through funnels which are located inside the
magnetosphere.
   They penetrate inwards through
the Rayleigh-Taylor type instability
   (e.g. Arons \& Lea 1976)
   up to some distance $r_{1}$,
and then form regular funnel streams along
the field lines.
   The distance of penetration
depends on the ratio $r_A/R_*$.
 At relatively small values $r_A/R_*$, the equatorial funnels may
penetrate almost to the surface of the star, as shown in
Figure 6.
 At larger values of $r_A/R_*$,
the funnels move inward only part of the way.
  Thus, in the case of a
weak magnetic field and/or high accretion rate, the magnetospheric
radius $r_A\approx (1-3 )R_*$, and a  small adiabatic indexes
$\gamma$, the gap will not be empty and planets will continue to
migrate inward unless the tidal interaction or some other force will
prevent them against falling to the star.
   In the opposite case of a larger
magnetosphere, the equatorial funnels will occupy only a part of the
gap and planet may survive longer inside the innermost gap.

\section{Conclusions}

We conclude that for  typical parameters of solar-type
protostars a very low-density magnetospheric gap forms.
   We show that the rate of inward migration of protoplanets
is greatly reduced within this gap.
   This gap may explain the
observed accumulation of planets
at the distances $\sim (0.04-0.05)$
AU from the star.
   Note however that there may
be significant matter density
within $r_A$ in  cases of high misalignment angles
$\Theta \gtrsim 45^\circ$ where the part of the  funnel stream is
located in the equatorial plane.
   Further,  for low values of the
adiabatic index, $\gamma=1.1$, and  small misalignment angles,
$\Theta \lesssim 5^\circ$, matter may accrete in the equatorial
plane due to instabilities.  However, the azimuthal velocity
of the matter within $r_A$ differs substantially from
the Keplerian angular velocity of the planet.
 This difference in the
velocities acts to slow the planet's migration.

\acknowledgments{We thank Drs. G.V. Ustyugova
and A.V. Koldoba for valuable contributions to
our MHD simulation codes.  Also we thank the referee
for thoughtfull comments.
This work was supported in part by NASA grant
NNG05GL49G and  by NSF grant AST-0507760.}

 \end{document}